\documentclass[letterpaper,twocolumn,prl,aps,superscriptaddress,amsmath,amssymb,floatfix]{revtex4}
\usepackage{mathptmx}
\usepackage[latin9]{inputenc}
\usepackage{color}
\usepackage{amsmath}
\usepackage{amssymb}
\usepackage{graphicx}
\usepackage{esint}
\usepackage[unicode=true,pdfusetitle,
 bookmarks=true,bookmarksnumbered=false,bookmarksopen=false,
 breaklinks=false,pdfborder={0 0 1},backref=false,colorlinks=true]
 {hyperref}
\hypersetup{
 pdfborderstyle=,colorlinks,linkcolor=blue,citecolor=blue}

\makeatletter

\usepackage{mathrsfs}

\usepackage{textcomp}
\usepackage{epstopdf}
\usepackage{verbatim}
\pdfpageheight\paperheight
\pdfpagewidth\paperwidth

\@ifundefined{textcolor}{}{%
 \definecolor{BLACK}{gray}{0}
 \definecolor{WHITE}{gray}{1}
 \definecolor{RED}{rgb}{1,0,0}
 \definecolor{GREEN}{rgb}{0,1,0}
 \definecolor{BLUE}{rgb}{0,0,1}
 \definecolor{CYAN}{cmyk}{1,0,0,0}
 \definecolor{MAGENTA}{cmyk}{0,1,0,0}
 \definecolor{YELLOW}{cmyk}{0,0,1,0}
}
\usepackage{xcolor}\usepackage{soul}
\newcommand{\bra}[1]{\ensuremath{\left\langle#1\right|}}
\newcommand{\ket}[1]{\ensuremath{\left|#1\right\rangle}}
\definecolor{blue}{rgb}{0,0,1}
\definecolor{red}{rgb}{1,0,0}
\definecolor{green}{rgb}{0,1,0}

\makeatother

\begin{document}
\title{Experimental simulation of open quantum system dynamics via Trotterization}
\author{J.~Han}
\thanks{These two authors contributed equally to this work.}
\affiliation{Center for Quantum Information, Institute for Interdisciplinary Information
Sciences, Tsinghua University, Beijing 100084, China}
\author{W.~Cai}
\thanks{These two authors contributed equally to this work.}
\affiliation{Center for Quantum Information, Institute for Interdisciplinary Information
Sciences, Tsinghua University, Beijing 100084, China}
\author{L.~Hu}
\affiliation{Center for Quantum Information, Institute for Interdisciplinary Information
Sciences, Tsinghua University, Beijing 100084, China}
\author{X.~Mu}
\affiliation{Center for Quantum Information, Institute for Interdisciplinary Information
Sciences, Tsinghua University, Beijing 100084, China}
\author{Y.~Ma}
\affiliation{Center for Quantum Information, Institute for Interdisciplinary Information
Sciences, Tsinghua University, Beijing 100084, China}
\author{Y.~Xu}
\affiliation{Center for Quantum Information, Institute for Interdisciplinary Information
Sciences, Tsinghua University, Beijing 100084, China}
\author{W.~Wang}
\affiliation{Center for Quantum Information, Institute for Interdisciplinary Information
Sciences, Tsinghua University, Beijing 100084, China}
\author{H.~Wang}
\affiliation{Center for Quantum Information, Institute for Interdisciplinary Information
Sciences, Tsinghua University, Beijing 100084, China}
\author{Y.~P.~Song}
\affiliation{Center for Quantum Information, Institute for Interdisciplinary Information
Sciences, Tsinghua University, Beijing 100084, China}
\author{C.-L.~Zou}
\email{clzou321@ustc.edu.cn}

\affiliation{CAS Key Laboratory of Quantum Information, University of Science and
Technology of China, Hefei, Anhui 230026, China}
\author{L.~Sun}
\email{luyansun@tsinghua.edu.cn}

\affiliation{Center for Quantum Information, Institute for Interdisciplinary Information
Sciences, Tsinghua University, Beijing 100084, China}
\begin{abstract}
Digital quantum simulators provide a diversified tool for solving
the evolution of quantum systems with complicated Hamiltonians and
hold great potential for a wide range of applications. Although much attention is paid to the unitary evolution of closed quantum
systems, dissipation and noise are vital in understanding the
dynamics of practical quantum systems. In this work, we experimentally demonstrate a digital simulation of an open quantum system in a controllable Markovian environment with the assistance of a single ancillary qubit. By Trotterizing the quantum Liouvillians, the continuous evolution of an open quantum system is effectively realized, and its application in error mitigation
is demonstrated by adjusting the simulated noise intensities. High-order
Trotter for open quantum dynamics is also experimentally investigated
and shows higher accuracy. Our results represent a significant step towards hardware-efficient simulation of open quantum systems and error mitigation in quantum algorithms in noisy intermediate-scale quantum systems.
\end{abstract}
\maketitle
%
Quantum computers and quantum simulators attract great attentions for their unprecedented capability in information processing tasks, such as executing quantum algorithms for universal computation~\cite{Grover1997,Bacon2010} and solving dynamics of many-body quantum systems~\cite{Feynman1982Simulating,Buluta108}. Over the past two decades, great efforts have been dedicated to building quantum machines for implementing these tasks~\cite{Altman2019,Alexeev2019}, and exciting progress has been achieved. Recently, quantum processors composed of a few dozens of qubits have been realized, and the demonstration of ``quantum supremacy"~\cite{Google2019} proves the advantage of quantum computing.

However, quantum information is vulnerable to the noisy environment. The behavior of practical quantum systems deviates from the ideal model and their evolution follows non-unitary quantum dynamics~\cite{Nielsen2010,Koch2016}. Therefore, applications of noisy intermediate-scale quantum (NISQ)~\cite{Preskill2018} devices are limited. Quantum error correction techniques and fault-tolerant quantum architectures have been proposed to overcome this obstacle~\cite{Terhal2015Quantum,Devitt_2013,Terhal2017,2016arXiv161003507G}. However, they demand quantum hardware with high performance and consume a great amount of quantum resources, making them impractical with current technologies. Instead of eliminating noise, one can also synthesize noise to explore the potential of NISQ devices. For example, error mitigation (EM) has recently been proposed as a promising approach to improve the accuracy of quantum information processing by varying noise intensity and extrapolating the results from a collection of experiments~\cite{Temme2016Error,LiPRX2017,Kandala2019}. Furthermore, quantum noise also plays a significant role for the simulation of the non-equilibrium phase transitions~\cite{DallaTorre2010}, driven-dissipative phase transitions~\cite{Schindler2013,Heugel2019}, and non-Hermitian topological phenomena~\cite{Kawabata2019} that appear in quantum many-body systems. Therefore, the ability to simulate an environment with controllable noise intensity is of fundamental importance.

\begin{figure}
	\includegraphics{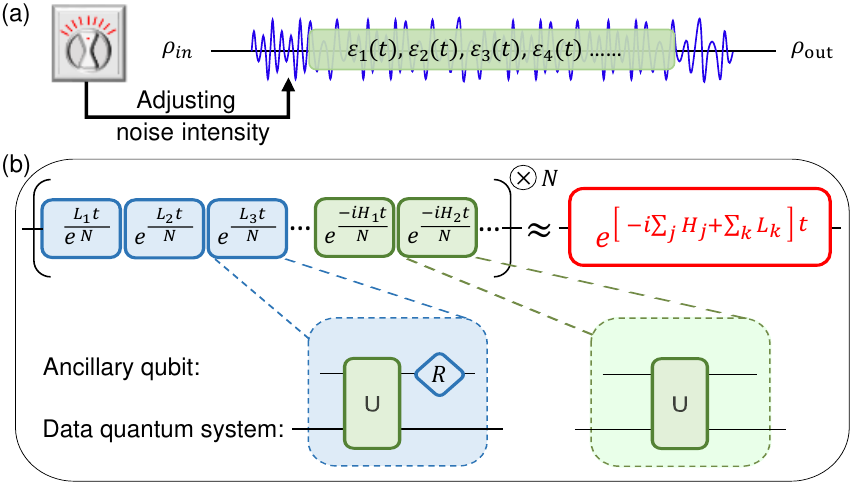} \caption{(\textbf{a})
		The schematic of open quantum system evolution with a controllable
		environment. (\textbf{b}) The Trotter scheme for simulating open quantum
		system dynamics, which consists of non-unitary jump operators $L_{k}$
		and coherent unitary evolution due to Hamiltonian $H_{j}$. The jump
		operators can be realized efficiently by a unitary gate on the combined
		system of the data quantum system and an ancillary qubit followed
		by resetting the ancillary qubit.}
	\label{fig:Fig1} \vspace{-6pt}
\end{figure}

\begin{figure*}
\includegraphics{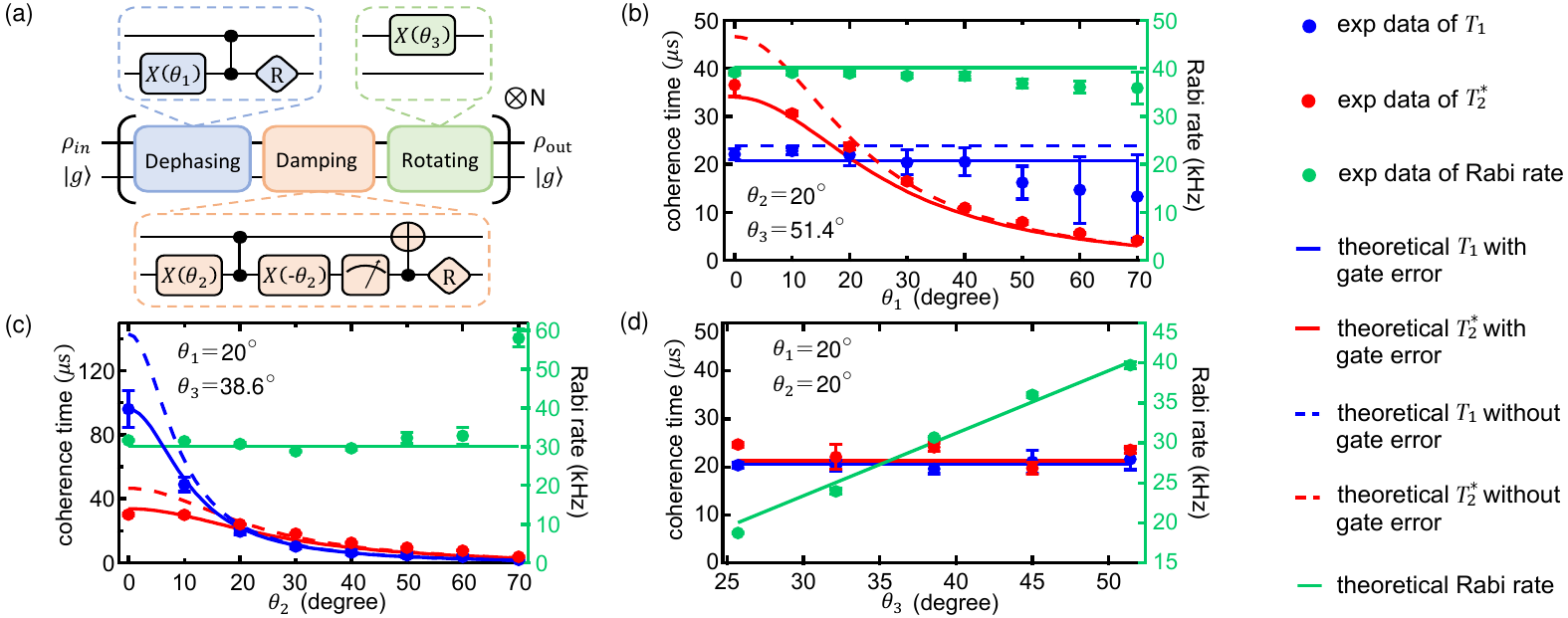} \caption{ {} (\textbf{a}) Experimental sequence for simulating the continuous evolution of an open quantum system by repetitively
implementing dephasing, damping and rotating Liouvillians. The detailed
schemes for realizing these Liouvillians with the assistance of an
ancillary qubit are shown in the dashed boxes. (\textbf{b})-(\textbf{d})
Experimentally extracted results of energy relaxation time $T_{1}$
(blue), phase relaxation time (Ramsey time) $T_{2}^{*}$ (red), and
Rabi oscillation rate $\Omega$ (green) for varying control parameters
$\theta_{1}$, $\theta_{2}$, and $\Omega$ of the Trotterized Liouvillians.
For all experiments, the Trotter step number $N=13$ and, except for the variable, the other two parameters are fixed: $\theta_{2}=20^{\circ}$, $\theta_{3}=51.4^{\circ}$ for (\textbf{b}); $\theta_{1}=20^{\circ}$, $\theta_{3}=38.6^{\circ}$ for (\textbf{c}); $\theta_{1}=20^{\circ}$, $\theta_{2}=20^{\circ}$ for (\textbf{d}).}
\label{fig:Fig2} \vspace{-6pt}	
\end{figure*}


In this work, we propose and demonstrate a digital simulation of the
continuous evolution of an open quantum system based on a Trotterization approach in a superconducting circuit. The Trotterization of the open quantum system dynamics is implemented repetitively by resetting and coupling an ancilla qubit to the data quantum system. We realize precise intensity adjustments for different types
of noise and present a proof-of-principle EM demonstration for estimating the zero-noise dephasing rate of the data system. Additionally, a 2nd-order Trotterization is applied to the system and shown to have higher precision. The approach to simulating
open quantum system dynamics demonstrated in this work would help
to understand and control quantum noise~\cite{DeVega2017}, realize universal digital quantum simulators~\cite{Lloyd1996}, and pave the way towards practical NISQ technologies~\cite{Preskill2018}.

%
The quantum dynamics of an open system {[}Fig.~\ref{fig:Fig1}(a){]}
can be described by the master equation~\cite{Gardiner2004Quantum,Breuer2002},
which can be written most generally in the Lindblad form~\cite{Lindblad1976On}
as $d\rho/dt=\sum_{j}\mathfrak{L}_{j}(\rho)+\sum_{k}\mathcal{L}_{k}(\rho)$,
where $\rho$ denotes the density operator of the system, the coherent
Liouvillian $\mathfrak{L}_{j}(\rho)=-i[H_{j},\rho]$ represents the
evolution due to the $j$-th Hamiltonian component $H_{j}$, and the incoherent Liouvillian $\mathcal{L}_{k}(\rho)=2L_{k}\rho L_{k}^{\dagger}-\{L_{k}^{\dagger}L_{k},\rho\}$
describes the quantum jump evolution due to the $k$-th jump operator
$L_{k}$. Thus, the evolution of the open system can be described
as $\rho(t)=e^{(\sum_{j}{\mathfrak{L}_{j}}+\sum_{k}{\mathcal{L}_{k}})t}\rho(0)$.

For the unitary evolution of a closed quantum system, the implementation
of the Hamiltonian $H=\sum H_{j}$ could be approximated by the Trotter
approach, i.e. the target unitary evolution with a duration $t$ is
discretized into $N$ steps and each step is decomposed into unitary
gates $e^{iH_{j}t/N}$ that are implemented alternatively~\cite{Lloyd1996,Trotter1959On,Suzuki1993Improved}: $\lim_{N\to\infty}{\left(\prod_{j}e^{iH_{j}t/N}\right)^{N}}=e^{iHt}.$
Such a digital approach has been widely adopted in quantum simulations
with complicated Hamiltonians both experimentally and theoretically~\cite{Lanyon2011,Salathe2015,Barends2015,Heyl2019}. The generalization of Trotterization to the open quantum system shares the same spirit as that for a closed system
\begin{equation}
\lim_{N\to\infty}{\left(\prod_{j}e^{\mathfrak{L}_{j}t/N}\prod_{k}e^{\mathcal{L}_{k}t/N}\right)^{N}}=e^{(\sum_{j}{\mathfrak{L}_{j}}+\sum_{k}{\mathcal{L}_{k}})t}.\label{Trotter}
\end{equation}
It is worth noting that there are different choices in the ordering of Liouvillian superoperators and some (such as the 2nd-order Trotterization
demonstrated in the following) might have a better precision for finite $N$. The deviation
of the discretized temporal evolution from the target quantum evolution can be suppressed to $\mathcal{O}\left[\left(t/N\right)^{m+1}\right]$
for the $m$-th order Trotterization (see Ref.~\cite{Supplement}\nocite{Khaneja2005,DeFouquieres2011,Richardson1927The,Sidi2003Practical}). Each Trotter step consists of an incoherent elementary process $e^{\mathcal{L}_{j}t/N}$ corresponding
to a quantum channel~\cite{Nielsen2010,Hu2018} that realizes a linear completely-positive trace-preserving mapping, and a coherent
elementary process $e^{\mathfrak{L}_{k}t/N}$ that is a unitary gate. Therefore, this scheme of digital quantum simulation of open quantum systems could be carried
out by experimentally implementing elementary channels on the data quantum
system, which can be efficiently
implemented by discarding the ancilla as the environment after a unitary
gate on the composite ancilla-data quantum system (Fig.~\ref{fig:Fig1}b).

\begin{figure}
	\includegraphics{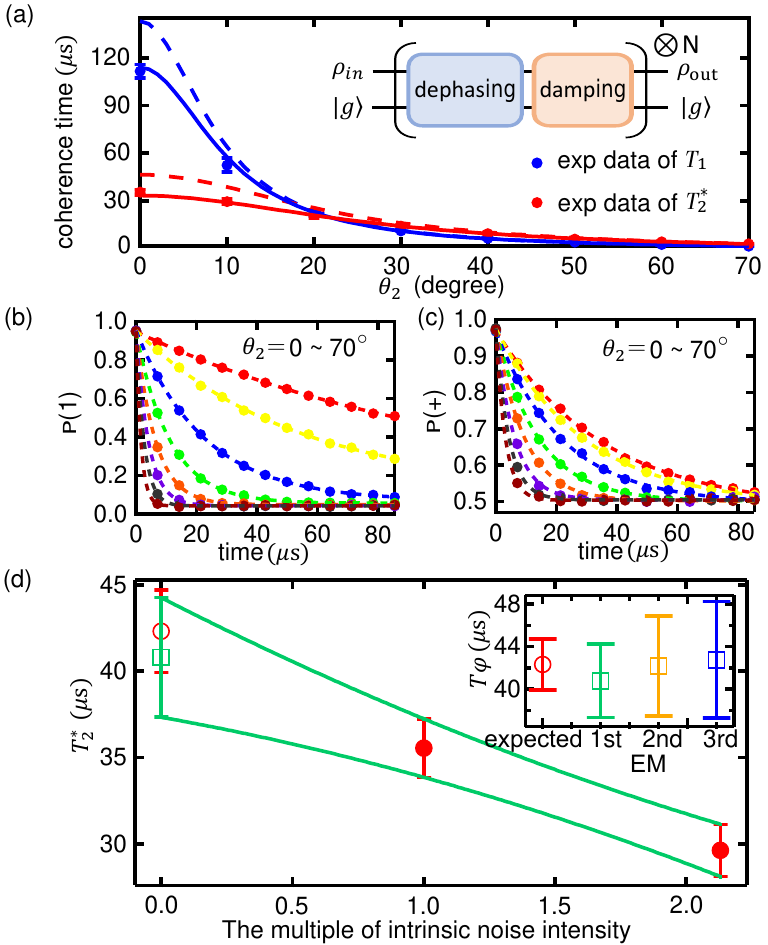} \caption{(\textbf{a}) Experimentally extracted $T_{1}$
		(blue) and $T_{2}^{*}$ (red) with varying $\theta_{2}$ while keeping
		$\theta_{1}=20^{\circ}$ in the Trotter scheme with repeated dephasing
		and damping channels only (the inset). Dots are derived results from
		experimental data, dashed lines are theoretically expected results
		from an ideal model, and solid lines are numerical results including
		gate errors. (\textbf{b}) and (\textbf{c}) Experimental evolution
		curves to give the results in (\textbf{a}). $P(1)$ and $P(+)$ are the populations of the excited
		state and $\ket{+}=(\ket{0}+\ket{1})/\sqrt{2}$ state of the data
		qubit, respectively. The dots are experimental data and dash lines
		are fitted curves. The colors from red to dark red correspond to $\theta_{2}$ varying from 0 to 70 degrees, respectively. (\textbf{d}) Estimation of dephasing time $T_{\phi}$
		by EM. The red points are the measured $T_{2}^{*}$ {[}first
		two data points in (\textbf{a}){]} and the green line shows the first-order
		Richardson extrapolation based on these two points. The red hollow
		circle with error bar is the calculated $T_{\phi}$ from the first
		pair of $T_{1}$ and $T_{2}^{*}$ in (\textbf{a}) through $1/T_{\phi}=1/T_{2}^{*}-1/2T_{1}$.
		Inset: error mitigation (EM) with Richardson extrapolations.}
	\label{fig:Fig3} \vspace{-6pt}
\end{figure}

We verify the proposed Trotterization approach
in a circuit quantum electrodynamics architecture~\cite{Devoret2013,Gu2017Microwave,Krantz2019,Wallraff2004,Paik,Hu2019,Ma2020}: the first two Fock states $\ket{0}$ and $\ket{1}$ of a microwave cavity constitute the data qubit undergoing open system evolutions; a dispersively coupled transmon qubit serves as the ancilla qubit. The device parameters can be found in Ref.~\cite{Supplement}. Figure~\ref{fig:Fig2}(a) shows the experimental sequence for simulating
the continuous evolution of the data qubit in an environment with controllable
noise intensity. We mimic the dephasing (jump operator
$L=\frac{\sqrt{\gamma_{1}}}{2}\sigma_{z}$) and the damping ($L=\sqrt{\gamma_{2}}\sigma_{-}$)
environment for the data qubit under continuous on-resonance drive,
with the corresponding channels being realized with measurement-based
adaptive control~\cite{Hu2018}. Here, $\gamma_1$ and $\gamma_2$ are the jumping rates. We first initialize the data qubit
into a specific state $\rho_{in}$ and the ancilla to the ground state
$\ket{g}$, then repetitively implement dephasing, damping, and rotating
channels alternatively for $N$ times, and finally measure the output
state of the data qubit by mapping its information to the ancilla through a decoding unitary, followed by the ancilla measurement~\cite{Supplement}.

When repeating the quantum circuits [Fig.~\ref{fig:Fig2}(a)] with
a period of $\tau_{0}=3.56~\mathrm{\mu s}$, the equivalent
environment noise strength can be tuned by the dephasing parameter
$\theta_{1}$, corresponding to a dephasing rate $\gamma_{1}=-{\rm ln}([2{\rm cos}^{2}(\theta_{1}/2)-1])/\tau_{0}$,
and the damping parameter $\theta_{2}$, corresponding to a damping rate
$\gamma_{2}=-{\rm ln}({\rm cos}^{2}\theta_{2})/\tau_{0}$. Then, the
resulting coherence times of the data qubit are $1/{T_{1}}=-{\rm ln}({\rm cos}^{2}\theta_{2})/\tau_{0}+1/{{T_{1}}^{0}}$
and $\begin{aligned}1/{T_{2}}= & -{\rm ln}([2{\rm cos}^{2}(\theta_{1}/2)-1])/\tau_{0}-{\rm ln}({\rm cos}^{2}\theta_{2})/2\tau_{0}+1/{{T_{2}}^{0}}.\end{aligned}
$ Here, $T_{1}^{0}$ and $T_{2}^{0}$ are the intrinsic coherent times
of the data qubit. The on-resonance drive is implemented by the
rotating gate $X(\theta_{3})$, which is equivalent to a Rabi rate
$\Omega=\theta_{3}/{360^{\circ}}\tau_{0}$ on the data qubit. In
order to evaluate the simulation on the open quantum system dynamics,
we initialize $\rho_{in}$ into $\ket{0}$, $\left(\ket{0}+\ket{1}\right)/\sqrt{2}$,
$\left(\ket{0}+i\ket{1}\right)/\sqrt{2}$, and $\ket{1}$ separately,
and measure the corresponding evolution curves of these states on
the bases of the Pauli operators $\sigma_{x}$, $\sigma_{y}$, and
$\sigma_{z}$. Then, by a global fitting of the twelve evolution curves,
we obtain all experimental $T_{1}$, $T_{2}$, and $\Omega$,
and compare them with the theoretical expectations.

Figures~\ref{fig:Fig2}(b)-(d) illustrate the experimental results
for the data qubit with simulated dephasing, damping and
rotation channels; all acting simultaneously. We separately fix two parameters
and regulate the third one to verify the effectiveness, controllability,
and flexibility of each adjustment. The experimental results are fairly
consistent with the expected ones including the gate errors (solid
lines), but with discrepancies at large $\theta_{1,2}$. These deviations
mainly come from two aspects: the high-order effects of Trotterization
become significant for large $\theta_{1,2}$ and the fitting errors become
larger when the system has a faster decoherence rate for larger $\theta_{1,2}$.
This experiment confirms the effective simulation of a Markovian environment
for an open system evolution, and such an approach not only enables
the simulation of arbitrary noise, but also allows for a wide range
of tunable noise intensity.

The tunability of the environmental noise intensity allows the precise
control of open quantum system dynamics, and thus provides a valuable
tool to extrapolate the behavior of a quantum system with no presence
of the inevitable noise through the so-called EM method~\cite{Temme2016Error,LiPRX2017,Kandala2019}.
As a proof-of-concept experiment, we find the dephasing time $T_{\phi}$
in the Trotter scheme with repeated dephasing and damping channels
only [Fig.~\ref{fig:Fig3}(a) inset]. In this experiment, the
extra dephasing rate from the dephasing channel is fixed with $\theta_{1}=20^{\circ}$
while the damping rate in the damping channel through $\theta_{2}$
is varying. The measured $T_{1}$ and $T_{2}^{*}$
(Ramsey time) as a function of $\theta_{2}$ are shown in Fig.~\ref{fig:Fig3}(a).
Based on the first two data points of $T_{2}^{*}$, we can derive
$T_{\phi}=T_{2}^{*}$ for $T_{1}\to\infty$ by the 1st-order Richardson
extrapolations (see Ref.~\cite{Supplement}), as shown in Fig.~\ref{fig:Fig3}(d).
The resulting $T_{\phi}$ from this 1st-order extrapolation, as well
as the higher-order ones based on more $T_{2}^{*}$ points, agrees
well with the expected value {[}Fig.~\ref{fig:Fig3}(d) inset{]}.
This demonstration nevertheless shows the potential
of EM in more complicated simulations of a
quantum system coupled to an environment with digitized and tunable
noise intensity.


\begin{figure}
\includegraphics{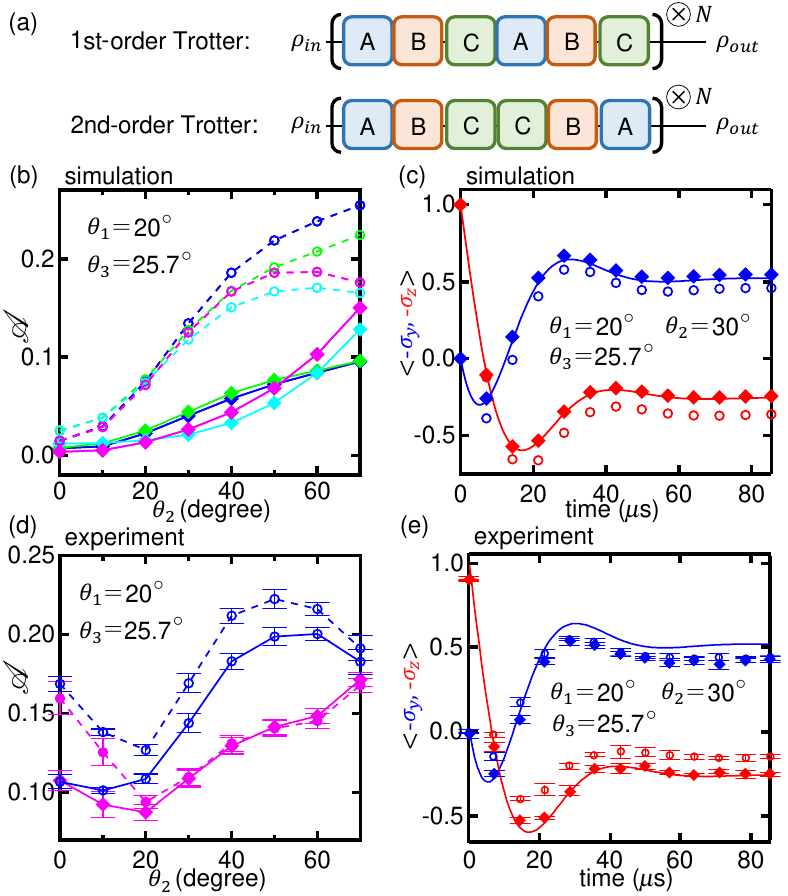} \caption{(\textbf{a}) Sequences of the 1st- and the 2nd-order Trotters for simulating
open quantum dynamics. (\textbf{b}) Numerical investigation of the accuracy $\mathcal{A}$ for six different operation permutations. We have different colors to denote different permutations of dephasing ($\mathcal{L}_{\mathrm{dph}}$), damping ($\mathcal{L}_{\mathrm{damp}}$)
and rotation ($\mathfrak{L}_{\mathrm{R}}$) in the form of $A-B-C$: blue, $\mathcal{L}_{\mathrm{dph}}-\mathcal{L}_{\mathrm{damp}}-\mathfrak{L}_{\mathrm{R}}$ and $\mathcal{L}_{\mathrm{damp}}-\mathcal{L}_{\mathrm{dph}}-\mathfrak{L}_{\mathrm{R}}$;
green, $\mathcal{L}_{\mathrm{damp}}-\mathfrak{L}_{\mathrm{R}}-\mathcal{L}_{\mathrm{dph}}$;
cyan, $\mathcal{L}_{\mathrm{dph}}-\mathfrak{L}_{\mathrm{R}}-\mathcal{L}_{\mathrm{damp}}$;
magenta, $\mathfrak{L}_{\mathrm{R}}-\mathcal{L}_{\mathrm{dph}}-\mathcal{L}_{\mathrm{damp}}$ and $\mathfrak{L}_{\mathrm{R}}-\mathcal{L}_{\mathrm{damp}}-\mathcal{L}_{\mathrm{dph}}$. Hollow and solid dots represent the 1st- and the 2nd-order Trotters, respectively, in (\textbf{b}-\textbf{e}).
(\textbf{c}) Numerically simulated evolutions of $\sigma_{y}$ (blue)
and $\sigma_{z}$ (red) with $\theta_{1}=20^{\circ}$, $\theta_{2}=30^{\circ}$, and
$\theta_{3}=25.7^{\circ}$. The solid lines are the target evolution
curves based on the master equation. (\textbf{d}) Experimental accuracy $\mathcal{A}$ for
the 1st-order Trotter (blue hollow dots, dashed line) with
permutation $\mathcal{L}_{\mathrm{dph}}-\mathcal{L}_{\mathrm{damp}}-\mathfrak{L}_{\mathrm{R}}$
and the 2nd-order Trotter (magenta solid dots, dashed line) with
permutation $\mathfrak{L}_{\mathrm{R}}-\mathcal{L}_{\mathrm{dph}}-\mathcal{L}_{\mathrm{damp}}$,
respectively. The solid lines correspond to the accuracy when
gate and measurement errors are considered in the target evolution as the reference. (\textbf{e}) Measured evolution curves of $\sigma_{y}$ (blue) and
$\sigma_{z}$ (red) with $\theta_{1}=20^{\circ}$, $\theta_{2}=30^{\circ}$, and
$\theta_{3}=25.7^{\circ}$. The solid lines are the same target evolution curves as in (\textbf{c}).}
\label{fig:Fig4} \vspace{-6pt}

\end{figure}

%
As observed in our experiments, the errors of Trotterization appear
larger for larger $\theta_{1,2}$ due to the finite Trotter step size
$\tau_{0}$. In principle, the error could be suppressed by reducing
$\tau_{0}$. However, the imperfection of the ancilla imposes a restriction
of $\tau_{0}$ (see Ref.~\cite{Supplement}), and the finite gate
time is also an intrinsic limitation. Therefore, we propose and demonstrate
the higher-order Trotter scheme for the open quantum system simulation
to achieve a better precision. The higher-order Trotter scheme is
constructed from the generalized Trotter scheme mentioned above by
symmetrization~\cite{Suzuki1976}, as illustrated in Fig.~\ref{fig:Fig4}(a).
For general Liouvillians $\mathbb{L}_{j}$ ($j\in\left\{ 1,..,m\right\} $)
that include all coherent and incoherent components, it can be proved that ~\cite{Kliesch2011,Supplement}
\begin{equation}
e^{\sum_{j=1}^{m}\mathbb{L}_{j}\Delta t}=\Pi_{j=1}^{m}e^{\mathbb{L}_{j}\Delta t/2}\Pi_{j=m}^{1}e^{\mathbb{L}_{j}\Delta t/2}+\mathcal{O}\left(\Delta t^{3}\right)
\end{equation}
for the 2nd-order Trotter formula with a precision of $\mathcal{O}(\Delta t^{3})$,
in comparison to
\begin{equation}
e^{\sum_{j=1}^{m}\mathbb{L}_{j}\Delta t}=\Pi_{j=1}^{m}e^{\mathbb{L}_{j}\Delta t}+\mathcal{O}\left(\Delta t^{2}\right)
\end{equation}
for the 1st-order Trotter.

Different permutations of the Liouvillians
$\left\{ \mathbb{L}_{j}\right\} $ in the Trotterization can have different accuracies. This is straightforward to understand for the case
with dephasing ($\mathcal{L}_{\mathrm{dph}}$), damping ($\mathcal{L}_{\mathrm{damp}}$),
and rotating ($\mathfrak{L}_{\mathrm{R}}$) channels
{[}Fig.~\ref{fig:Fig2}(a){]}, because $\mathfrak{L}_{\mathrm{R}}$
is not commutative to either $\mathcal{L}_{\mathrm{dph}}$ or $\mathcal{L}_{\mathrm{damp}}$. As $N$ is finite in the
practical experiment, the incommutability between channels leads to
the differences in the simulation accuracy for different channel permutations.

To quantify the precision of the quantum dynamics simulation, we define
the accuracy as $\mathcal{A}=\sqrt{\sum_{x,j}(x_{j}-x_{j}^{0})^{2}}/\sqrt{N}$,
where $x\in\left\{ \left\langle \sigma_{x}\right\rangle ,\left\langle \sigma_{y}\right\rangle ,\left\langle \sigma_{z}\right\rangle \right\} $
are the expectation values, $j=1,2,...,N$ denotes the step of the
evolution, and $x_j^{0}$ are the expectations of the target evolution.

We first numerically simulate the system evolutions as a function of
$\theta_{2}$ for different permutations of Liouvillians ($\mathcal{L}_{\mathrm{dph}}$,
$\mathcal{L}_{\mathrm{damp}}$, and $\mathfrak{L}_{\mathrm{R}}$) based on the Trotter scheme also with the intrinsic coherence times of the data qubit, and calculate their accuracies. For an initial state $\rho_{in}=\ket{1}\bra{1}$, the results are summarized in Fig.~\ref{fig:Fig4}(b). Different permutations
indeed have different accuracies and the commutative relation leads to the coincidence
between the sequences $\mathcal{L}_{\mathrm{dph}}-\mathcal{L}_{\mathrm{damp}}-\mathfrak{L}_{\mathrm{R}}$
and $\mathcal{L}_{\mathrm{damp}}-\mathcal{L}_{\mathrm{dph}}-\mathfrak{L}_{\mathrm{R}}$,
as well as between $\mathfrak{L}_{\mathrm{R}}-\mathcal{L}_{\mathrm{dph}}-\mathcal{L}_{\mathrm{damp}}$
and $\mathfrak{L}_{\mathrm{R}}-\mathcal{L}_{\mathrm{damp}}-\mathcal{L}_{\mathrm{dph}}$.
More importantly, we find that the 2nd-order Trotter outperforms the
1st-order Trotter for all possible sequences and confirm the validity
of high-order Trotterization in simulating open quantum dynamics.

In Fig.~\ref{fig:Fig4}(c), numerically simulated temporal evolution curves for $\sigma_{y}$ and $\sigma_{z}$ of the data qubit are provided with $\theta_{1}=20^{\circ}$, $\theta_{2}=30^{\circ}$, and $\theta_{3}=25.7^{\circ}$. The 2nd-order Trotter shows better consistence with the target evolution (solid lines) than the 1st-order Trotter. It is worth noting that the total number of Trotter
steps is fixed when comparing the 1st- and the 2nd-order Trotters {[}as
indicated in Fig.~\ref{fig:Fig4}(a){]}, and the results unambiguously
reveal the advantage of the higher-order Trotterization in terms of
the higher accuracy by re-ordering.


We then perform more realistic numerical simulations of $\mathcal{A}$ based on QuTip~\cite{Johansson2012,Johansson2013} with the experimental pulses and the intrinsic coherence times of the data qubit (see Ref.~\cite{Supplement}), and obtain the optimal permutations for the 1st-order Trotter ($\mathcal{L}_{\mathrm{dph}}-\mathcal{L}_{\mathrm{damp}}-\mathfrak{L}_{\mathrm{R}}$) and the 2nd-order Trotter ($\mathfrak{L}_{\mathrm{R}}-\mathcal{L}_{\mathrm{dph}}-\mathcal{L}_{\mathrm{damp}}$),
respectively. Figure~\ref{fig:Fig4}(d) shows the corresponding experimental results of the accuracies as a function of $\theta_2$. Clearly, the 2nd-order Trotter always has a significant higher accuracy, especially when $\theta_{2}$ is
large. By taking into account the gate and measurement errors, which modify the effective coherence times in the master equation for the target evolution, the variation tendencies of the adjusted accuracies (solid lines) are also consistent with the numerical simulations in Fig.~\ref{fig:Fig4}(b). Figure~\ref{fig:Fig4}(e) shows the experimentally measured evolutions of the system under the simulated open quantum dynamics, in which the 2nd-order Trotter (solid
dots) show a better agreement with the target evolution (solid lines) than
the 1st-order Trotter (hollow dots), as expected.

Although the data quantum system in current experiment is a two-level qubit, the extension to multi-qubit systems and high-dimensional qudit systems can be straightforward. It is generally considered that to realize an elementary channel of a quantum system that could be decomposed into two Kraus operators requires a two-dimensional ancillary system for unitary dilation~\cite{Robinson2018,Sparrow2018}. Furthermore, other theoretical works~\cite{Lloyd2001PRA,Shen2017PRB} have proved that arbitrary elementary quantum channels with more than two Kraus operators can be efficiently implemented in arbitrarily high-dimensional quantum systems ($d$ dimensions) with the assistance of only one ancilla qubit and at most $2\mathrm{log}_2d$ steps of adaptive control. Recently, with the same experimental setup, arbitrary elementary quantum channels on a high-dimensional data quantum system have been experimentally demonstrated~\cite{Cai2021}. Therefore, the Trotterization scheme for the universal simulation of open quantum system dynamics with high dimensions can be applied in the same way, holding the advantage of mimicking the quantum system in an environment with tunable noise intensity in a wide range.

In conclusion, a universal digital quantum simulator based on the
Trotterization scheme is demonstrated for studying the dynamics of
a two-level open quantum system. We also demonstrate the 2nd-order Trotter scheme for a more accurate simulation of the quantum evolution with a given step size.
Benefiting from the universality and tunability, interesting open quantum system phenomena could be explored and the EM technique can be implemented in NISQ-based digital quantum simulators.

\begin{acknowledgments}
\textbf{Acknowledgments}. This work was supported by National Key Research and Development Program of China (Grant No.2017YFA0304303), Key-Area Research and Development Program of Guangdong Provice (Grant No.2020B0303030001), the 	National Natural Science Foundation of China (Grant No.11925404 and 11874235), and a Grant (No.2019GQG1024) from the Institute for Guo Qiang, Tsinghua University. C.-L.Z. was supported by National Natural Science Foundation of China (Grant No.11874342, 11922411, and 12061131011) and Anhui Initiative in Quantum Information Technologies (AHY130200).
\end{acknowledgments}

%

%

%

\end{document}


\onecolumngrid
\renewcommand{\thefigure}{S\arabic{figure}}
\setcounter{figure}{0}
\renewcommand{\thepage}{S\arabic{page}}
\setcounter{page}{1}
\renewcommand{\theequation}{S.\arabic{equation}}
\setcounter{equation}{0}
\setcounter{section}{0}
\begin{center}
{\Large{}SUPPLEMENTARY MATERIALS FOR}{\Large\par}
\par\end{center}
\title{Experimental simulation of open quantum system dynamics via Trotterization}
\author{J.~Han}
\thanks{These two authors contributed equally to this work.}
\affiliation{Center for Quantum Information, Institute for Interdisciplinary Information
Sciences, Tsinghua University, Beijing 100084, China}
\author{W.~Cai}
\thanks{These two authors contributed equally to this work.}
\affiliation{Center for Quantum Information, Institute for Interdisciplinary Information
Sciences, Tsinghua University, Beijing 100084, China}
\author{L.~Hu}
\affiliation{Center for Quantum Information, Institute for Interdisciplinary Information
Sciences, Tsinghua University, Beijing 100084, China}
\author{X.~Mu}
\affiliation{Center for Quantum Information, Institute for Interdisciplinary Information
Sciences, Tsinghua University, Beijing 100084, China}
\author{Y.~Ma}
\affiliation{Center for Quantum Information, Institute for Interdisciplinary Information
Sciences, Tsinghua University, Beijing 100084, China}
\author{Y.~Xu}
\affiliation{Center for Quantum Information, Institute for Interdisciplinary Information
Sciences, Tsinghua University, Beijing 100084, China}
\author{W.~Wang}
\affiliation{Center for Quantum Information, Institute for Interdisciplinary Information
Sciences, Tsinghua University, Beijing 100084, China}
\author{H.~Wang}
\affiliation{Center for Quantum Information, Institute for Interdisciplinary Information
Sciences, Tsinghua University, Beijing 100084, China}
\author{Y.~P.~Song}
\affiliation{Center for Quantum Information, Institute for Interdisciplinary Information
Sciences, Tsinghua University, Beijing 100084, China}
\author{C.-L.~Zou}
\email{clzou321@ustc.edu.cn}

\affiliation{CAS Key Laboratory of Quantum Information, University of Science and
Technology of China, Hefei, Anhui 230026, China}
\author{L.~Sun}
\email{luyansun@tsinghua.edu.cn}

\affiliation{Center for Quantum Information, Institute for Interdisciplinary Information
Sciences, Tsinghua University, Beijing 100084, China}

\maketitle
\tableofcontents{}

\section{Theory of open quantum system dynamics}
\begin{figure}[b]
\centering \includegraphics[width=8cm]{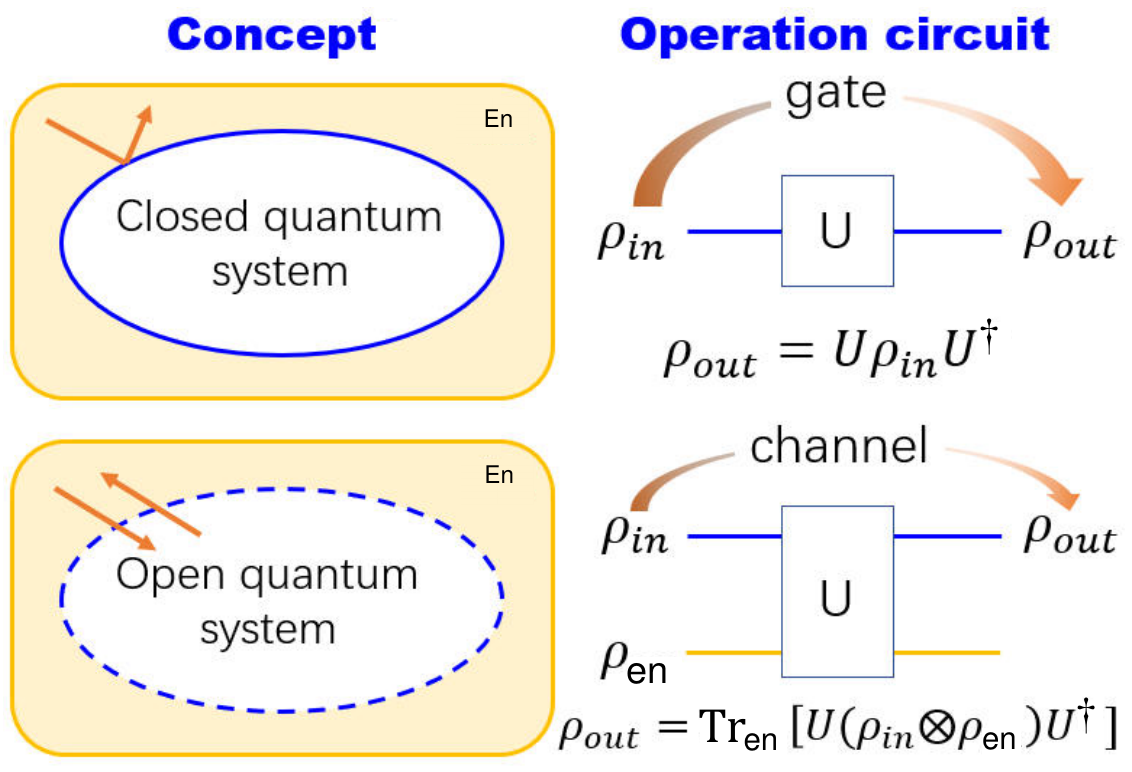} \caption{The conceptual illustration of closed and open quantum systems.}
\label{fig:concept_open_system}
\end{figure}

As shown in Fig.~\ref{fig:concept_open_system}, a closed quantum
system is an ideal quantum system that has no interaction with the
environment. The evolution of a closed quantum system can be represented
by a unitary gate acting on the quantum state of the system. For an open quantum system, however, there are interactions between the quantum
system and the environment, and thus the evolution of the quantum system is no longer unitary due to these couplings.

\subsection{Quantum Master Equation}
The continuous evolution of a Markovian open quantum system can be
described by the master equation~\cite{Breuer2002,Gardiner2004Quantum},
which is a differential equation to properly describe the non-unitary behavior
of the system. The master equation can be written most generally in the Lindblad
form~\cite{Lindblad1976On} as
\begin{equation}
\begin{aligned}\frac{d\rho}{dt} & =-i\sum_{j}[H_{j},\rho]+\sum_{k}[2L_{k}\rho L_{k}^{+}-\{L_{k}^{+}L_{k},\rho\}]\\
 & =\sum_{j}\mathfrak{L}_{j}(\rho)+\sum_{k}\mathcal{L}_{k}(\rho),\label{master_{e}quation}
\end{aligned}
\end{equation}
where $H_{j}$ are the Hamiltonians, which are Hermitian operators
representing the coherent part of the dynamics, and $L_{k}$ are the
Lindblad jump operators representing the coupling of the system to
the environment. With Liouvillian $\mathfrak{L}_{j}(\rho)$ representing
the coherent evolution and $\mathcal{L}_{k}(\rho)$ describing the incoherent
evolution, the dynamics of the system can be described as
\begin{equation}
\rho(t)=e^{(\sum_{j}{\mathfrak{L}_{j}}+\sum_{k}{\mathcal{L}_{k}})t}\rho(0).\label{Liouvillian_trotter}
\end{equation}

For example, the dephasing of a qubit could be described by
\begin{equation}
\frac{d\rho}{dt}=-i[H,\rho]+\frac{\gamma_{1}}{2}\left(\sigma_{z}\rho\sigma_{z}-\rho\right),\label{master_equation-dephasing}
\end{equation}
with $\gamma_{1}$ being the dephasing rate. For a two-level atom
coupling to the vacuum and undergoing spontaneous emission, the master
equation to describe this process is
\begin{equation}
\frac{d\rho}{dt}=-i[H,\rho]+\gamma_{2}\left(2\sigma_{-}\rho\sigma_{+}-\sigma_{+}\sigma_{-}\rho-\rho\sigma_{+}\sigma_{-}\right),\label{master_equation-damping}
\end{equation}
where $H=\omega\sigma_{z}/2$, Lindblad operator $\sqrt{\gamma_{2}}\sigma_{-}$
represents the emission, and $\gamma_{2}$ is the spontaneous emission rate.

\subsection{Quantum channel}
For an open quantum system (the data quantum system), the evolution could be described by a unitary on a larger closed system consisting of the data quantum system
and the environment. Then, the evolution from the input state $\rho_{in}$
to the output state $\rho_{out}$ of the data quantum system could
be obtained by discarding the environment as
\begin{equation}
\rho_{out}=\mathrm{Tr}_{\mathrm{en}}\left[U(\rho_{in}\otimes\rho_{\mathrm{en}})U^{\dagger}\right],
\end{equation}
with $\mathrm{Tr}_{en}$ denoting the partial trace of the environmental
degrees of freedom. Such a mapping could describe either a temporal
evolution or travelling through a distance of the quantum system,
and is also called ``quantum channel''~\cite{Nielsen2010,Breuer2002}.
A quantum channel can be represented as a sum of the Kraus operators~\cite{Nielsen2010}
\begin{equation}
\mathcal{E}(\rho)=\sum_{k=1}^{m}E_{k}\rho E_{k}^{\dagger},
\end{equation}
where $\sum_{k=1}^{m}E_{k}^{\dagger}E_{k}=I$. As a special case,
if an irreducible representation of a channel has only one Kraus
operator, then the channel reduces to a unitary evolution:
\begin{equation}
\mathcal{E}(\rho)=U\rho U^{\dagger}.
\end{equation}

There are two quantum channels for the data qubit in our experiments.
The first one is the dephasing channel, which describes the loss of phase
information without losing energy. We can interpret the dephasing
channel as random rotations $R_{z}^{\theta}$ (a rotation of $\theta$ along the $z$-axis of the Bloch sphere) on the quantum state, the effect of which makes the off-diagonal elements of $\rho$ decay exponentially to $0$. According to Eq.~(\ref{master_equation-dephasing}), for a dephasing rate $\gamma_{1}$ and an evolution duration time $\tau$, the Kraus operators of a dephasing channel are:
\begin{equation}
\begin{aligned}E_{0} & =\left[\begin{array}{cc}
1 & 0\\
0 & e^{-\gamma_{1}\tau/2}
\end{array}\right],\\
E_{1} & =\left[\begin{array}{cc}
0 & 0\\
0 & \sqrt{1-e^{-\gamma_{1}\tau}}
\end{array}\right],
\end{aligned}
\end{equation}
where $1-e^{-\gamma_1\tau}$ represents the probability of the data qubit to dephase without
losing energy. Suppose an initial data qubit state is
\begin{equation}
\rho=\left[\begin{array}{cc}
a & b\\
b^{*} & c
\end{array}\right],\label{equ:initial_state_channel}
\end{equation}
then the output state of the data qubit after the dephasing channel is
\begin{equation}
\begin{aligned}\mathcal{E}_{\mathrm{dph}}(\rho) & =\sum_{k}E_{k}\rho E_{k}^{\dagger}\\
 & =\left[\begin{array}{cc}
a & be^{-\gamma_{1}\tau/2}\\
b^{*}e^{-\gamma_{1}\tau/2} & c
\end{array}\right].
\end{aligned}
\end{equation}

The second quantum channel used in our experiment is the amplitude
damping channel, which describes the effect of energy dissipation,
i.e. spontaneous emission. For example, an excited quantum state \ket{1}
always has a probability of losing its energy and finally returns to the ground state \ket{0}. For a damping rate $\gamma_{2}$ and an evolution duration
time $\tau$, the corresponding Kraus operators are
\begin{equation}
\begin{aligned}E_{0} & =\left[\begin{array}{cc}
1 & 0\\
0 & e^{-\gamma_{2}\tau/2}
\end{array}\right],\\
E_{1} & =\left[\begin{array}{cc}
0 & \sqrt{1-e^{-\gamma_{2}\tau}}\\
0 & 0
\end{array}\right].
\end{aligned}
\label{equation_elements_energy_dissipation_singlequbit}
\end{equation}
Suppose the initial state of a data qubit is described as Eq.~(\ref{equ:initial_state_channel}),
then the output state of this qubit after the amplitude damping channel
is
\begin{equation}
\begin{aligned}\mathcal{E}_{\mathrm{damp}}(\rho) & =\sum_{k}E_{k}\rho E_{k}^{\dagger}\\
 & =\left[\begin{array}{cc}
1-ce^{-\gamma_{2}\tau} & be^{-\gamma_{2}\tau/2}\\
b^{*}e^{-\gamma_{2}\tau/2} & ce^{-\gamma_{2}\tau}
\end{array}\right].
\end{aligned}
\end{equation}

\subsection{Trotterization of Liouvillians}
The achievable interaction Hamiltonian could be limited and localized
for a complex quantum system. Similar to the Trotterization of the Hamiltonian simulation~\cite{Trotter1959On,Suzuki1993Improved}, we could implement the Liouvillians of the system by alternatively implementing the local Liouvillians. For general Liouvillians $\mathbb{L}_{j}$ ($j\in\left\{ 1,..,m\right\} $)
that include both coherent and incoherent components, the complete
Liouvillian could be represented as $e^{\sum_{j=1}^{m}\mathbb{L}_{j}t}$.
The Trotter decomposition of the whole Liouvillian means: 1) divide
each component into $N$ pieces, i.e. the evolution duration for each
step is $\Delta t=\frac{t}{N}$; 2) implement each piece of each
channel in turn; 3) repeat the above sequence for $N$ times.
By Taylor expansion,
\begin{equation}
e^{\sum_{j=1}^{m}\mathbb{L}_{j}\Delta t}=I+\sum_{j=1}^{m}\mathcal{\mathbb{L}}_{j}\Delta t+\mathcal{O}(\Delta t^{2}).\label{Taylor_expansion_trotter}
\end{equation}
Here, we have the time normalized to the norms of the operators $\left\Vert \mathbb{L}_{j}\right\Vert $, and thus we have $\Delta t\ll1$ for $N\gg1$. Similarly, we have
\begin{equation}
\prod_{j}e^{\mathbb{L}_{j}\Delta t}=I+(\mathbb{L}_{1}+\mathbb{L}_{2}+...+\mathbb{L}_{m})\Delta t+\mathcal{O}(\Delta t^{2}).\label{Taylor_expansion_mul_trotter}
\end{equation}
Therefore,
\begin{equation}
e^{\sum_{j=1}^{m}\mathbb{L}_{j}\Delta t}=\prod_{j}e^{\mathbb{L}_{j}\Delta t}+\mathcal{O}(\Delta t^{2})
\end{equation}
and
\begin{align}
e^{\sum_{j=1}^{m}\mathbb{L}_{j}t} & =\left[e^{\sum_{j=1}^{m}\mathbb{L}_{j}\Delta t}\right]^{N}\\
 & =\left[\prod_{j}e^{\mathbb{L}_{j}\Delta t}+\mathcal{O}(\Delta t^{2})\right]^{N}\\
 & =\left(\prod_{j}e^{\mathbb{L}_{j}\Delta t}\right)^{N}+\sum_{k=1}^{N}C_{N}^{k}\left(\prod_{j}e^{\mathbb{L}_{j}\Delta t}\right)^{N-k}\mathcal{O}(\Delta t^{2k}).
\end{align}
Since $C_{N}^{k}<N^{k}$ and $\prod_{j}e^{\mathbb{L}_{j}\Delta t}=\mathcal{O}\left(1\right)$, we have $C_{N}^{k}\mathcal{O}(\Delta t^{2k})=C_{N}^{k}\mathcal{O}(\frac{1}{N^{2k}})<\mathcal{O}(\frac{1}{N^{k}})$. Therefore, $\sum_{k=1}^{N}C_{N}^{k}\left(\prod_{j}e^{\mathbb{L}_{j}\Delta t}\right)^{N-k}\mathcal{O}(\Delta t^{2k})=\mathcal{O}(\Delta t)$,
and we finally have
\begin{align}
e^{\sum_{j=1}^{m}\mathbb{L}_{j}t} & =\left(\prod_{j}e^{\mathbb{L}_{j}\frac{t}{N}}\right)^{N}+\mathcal{O}(\frac{1}{N}).
\label{eq:TrotterDecomposition}
\end{align}
By simply repeating the Liouvillians in a fixed order, the target
Liouvillians could in principle be realized with an imperfection $\mathcal{O}\left(1/N\right)$
for $N\rightarrow\infty$.

There is another way to get Eq.~(\ref{eq:TrotterDecomposition}). Based on the Baker-Campbell-Hausdorff (BCH) formula
\begin{align}
e^{\left(A+B\right)t} & =e^{At}e^{Bt}e^{-\frac{1}{2}[A,B]t^{2}}+\mathcal{O}\left(t^{3}\right)\label{eq:BCH1}\\
 &
 =e^{+\frac{1}{2}[A,B]t^{2}}e^{Bt}e^{At}+\mathcal{O}\left(t^{3}\right),
\label{eq:BCH2}
\end{align}
the Trotterization of Liouvillians could be error free if $\mathbb{L}_{j}$ is commutative with $\mathbb{L}_{k}$, $\forall k\neq j$. Therefore, the error is
closely related to the order of $\mathbb{L}_{j}$ in general. According to the BCH formula, we can get the accuracy of the 1st-order Trotter as
\begin{align}
e^{(\mathbb{L}_{1}+\mathbb{L}_{2})\Delta t} & =e^{\mathbb{L}_{1}\Delta t}e^{\mathbb{L}_{2}\Delta t}e^{-\frac{1}{2}[\mathbb{L}_{1}\Delta t,\mathbb{L}_{2}\Delta t]}+\mathcal{O}\left(\Delta t^{3}\right)\nonumber \\
 & =e^{\mathbb{L}_{1}\Delta t}e^{\mathcal{\mathbb{L}}_{2}\Delta t}\left(I+\mathcal{O}\left(\Delta t^{2}\right)\right)+\mathcal{O}(\Delta t^{3})\\
 & =e^{\mathbb{L}_{1}\Delta t}e^{\mathcal{\mathbb{L}}_{2}\Delta t}+\mathcal{O}(\Delta t^{2}).
\end{align}
Applying the iteration to $\left\{ \mathbb{L}_{j}\right\} $, we obtain
\begin{align}
e^{\sum_{j=1}^{m}\mathbb{L}_{j}\Delta t} & =\prod_{j}e^{\mathbb{L}_{j}\Delta t}+\mathcal{O}(\Delta t^{2}).
\end{align}

For the 2nd-order Trotter, we implement $2N$ Trotter steps with repetitions
of normal ordering $\left\{ \mathbb{L}_{1},\mathbb{L}_{2},...,\mathbb{L}_{j},...,\mathbb{L}_{m}\right\} $
followed by the reverse ordering $\left\{ \mathbb{L}_{m},...,\mathbb{L}_{j},...,\mathbb{L}_{2},\mathbb{L}_{1}\right\} $,
i.e.
\begin{equation}
\Pi_{j=1}^{m}e^{\mathbb{L}_{j}\Delta t/2}\Pi_{j=m}^{1}e^{\mathbb{L}_{j}\Delta t/2}.
\end{equation}
Because we have
\begin{equation}
e^{(\mathbb{L}_{1}+\mathbb{L}_{2})\Delta t}=e^{\frac{1}{2}(\mathbb{L}_{1}+\mathbb{L}_{2})\Delta t}e^{\frac{1}{2}(\mathbb{L}_{1}+\mathbb{L}_{2})\Delta t},
\label{Eq:2ndTrotter}
\end{equation}
by applying Eq.~(\ref{eq:BCH1}) and Eq.~(\ref{eq:BCH2}) to
the first term and the second term on the right side of Eq.~\ref{Eq:2ndTrotter}, respectively, we can get:
\begin{align}
 & e^{\frac{1}{2}(\mathbb{L}_{1}+\mathbb{L}_{2})\Delta t}e^{\frac{1}{2}(\mathbb{L}_{1}+\mathbb{L}_{2})\Delta t}\nonumber \\
= & \left[e^{\frac{1}{2}\mathbb{L}_{1}\Delta t}e^{\frac{1}{2}\mathbb{L}_{2}\Delta t}e^{-\frac{1}{4}[\mathbb{L}_{1}\Delta t,\mathbb{L}_{2}\Delta t]}+\mathcal{O}\left(\Delta t^{3}\right)\right]\nonumber \\
 & \times\left[e^{\frac{1}{4}[\mathbb{L}_{1}\Delta t,\mathbb{L}_{2}\Delta t]}e^{\frac{1}{2}\mathbb{L}_{2}\Delta t}e^{\frac{1}{2}\mathbb{L}_{1}\Delta t}+\mathcal{O}\left(\Delta t^{3}\right)\right]\\
= & e^{\frac{1}{2}\mathbb{L}_{1}\Delta t}e^{\frac{1}{2}\mathbb{L}_{2}\Delta t}e^{\frac{1}{2}\mathbb{L}_{2}\Delta t}e^{\frac{1}{2}\mathbb{L}_{1}\Delta t}+\mathcal{O}\left(\Delta t^{3}\right).
\end{align}
Applying the iteration to $\left\{ \mathbb{L}_{j}\right\}$, we then have:
\begin{equation}
e^{\sum_{j=1}^{m}\mathbb{L}_{j}\Delta t}=\Pi_{j=1}^{m}e^{\mathbb{L}_{j}\Delta t/2}\Pi_{j=m}^{1}e^{\mathbb{L}_{j}\Delta t/2}+\mathcal{O}\left(\Delta t^{3}\right).
\end{equation}
Therefore, we finally obtain
\begin{equation}
e^{\sum_{j=1}^{m}\mathbb{L}_{j}t}=\left(\Pi_{j=1}^{m}e^{\mathbb{L}_{j}\frac{t}{2N}}\Pi_{j=m}^{1}e^{\mathbb{L}_{j}\frac{t}{2N}}\right)^{N}+\mathcal{O}\left(\frac{1}{N^{2}}\right).
\end{equation}
This expression indicates that we can suppress the error from $\mathcal{O}\left(\frac{1}{N}\right)$
to $\mathcal{O}\left(\frac{1}{N^{2}}\right)$ by simply re-ordering
the Trotterization terms. This could be helpful in practical experiments
when $N$ is limited by the gate time or other factors.

\begin{figure*}
\includegraphics[width=14cm]{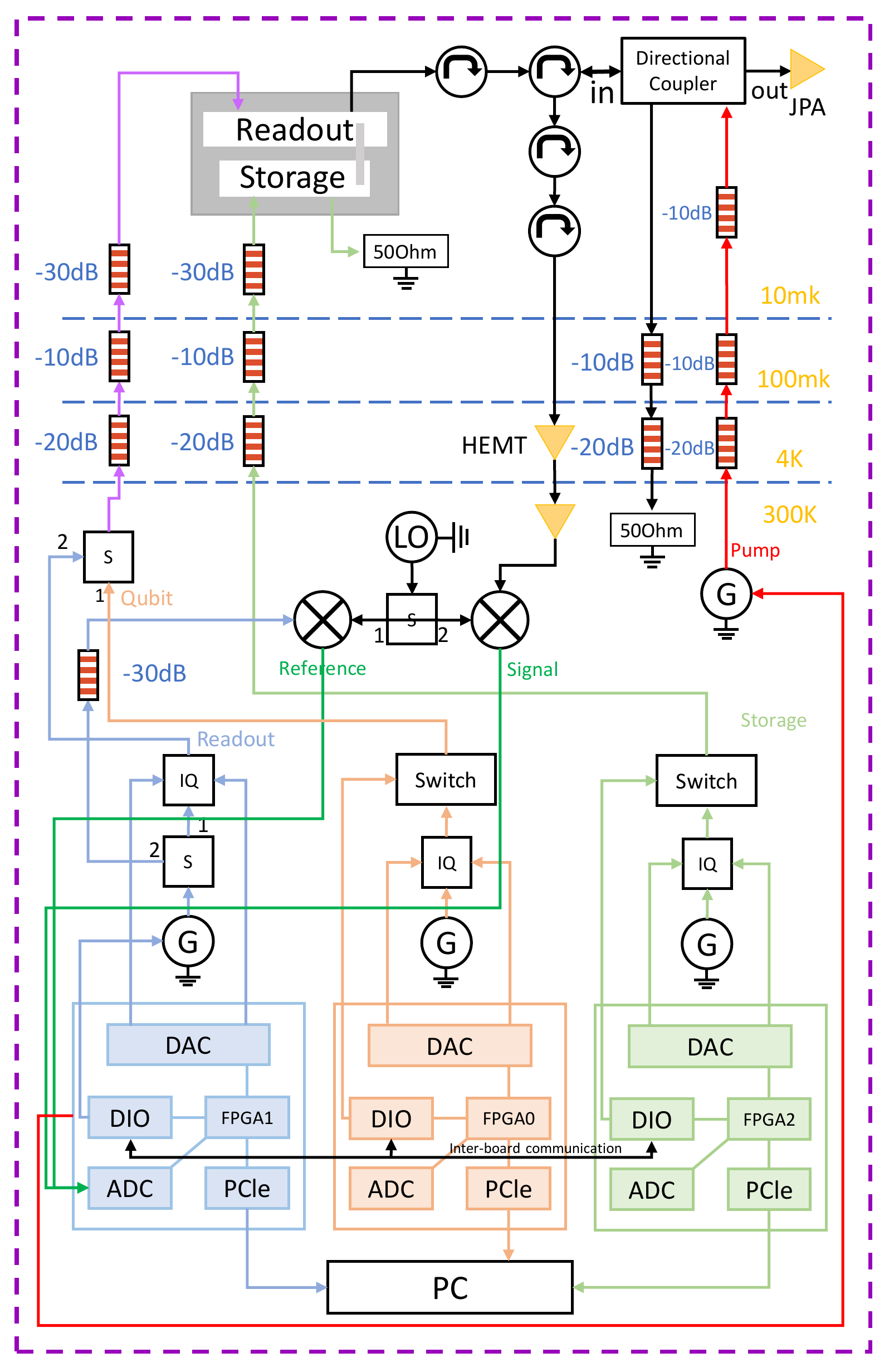} \caption{Schematic of the experimental setup.}
\label{fig:exp_diagram}
\end{figure*}

\section{Experiments}
\subsection{Experimental device and setup}
The experimental device consists of two superconducting waveguide cavity resonators and a transmon qubit dispersively coupled to them. One of the cavities (storage cavity) has a lifetime of $T_{1\mathrm{s}}=143~\mu$s and a Ramsey phase coherence time of $T_{2\mathrm{s}}=250~\mu$s, whose first two Fock states $\ket{0}$ and $\ket{1}$ constitute the two basis states of the data qubit, serving as the data quantum system undergoing open system evolutions. The transmon qubit, with a lifetime of $T_{1\mathrm{q}}=25~\mu$s, acts as an ancillary qubit to assist the control and readout of the data system. The other short-lived cavity (lifetime is $T_{1\mathrm{r}}=44$~ns) is used to readout the ancillary qubit with the help of a Josephson parametric amplifier for a high fidelity measurement. The readout of the data qubit is realized by mapping the information of the data qubit to the ancilla through a decoding unitary, followed by an ancilla measurement. The unitary is achieved with a gradient ascent pulse engineering (GRAPE)
technique~\cite{Khaneja2005,DeFouquieres2011} and transfers $(\alpha\ket{0}+\beta\ket{1})\ket{g}$ to $\ket{0}(\alpha\ket{g}+\beta\ket{e})$, where $\ket{g}$ and $\ket{e}$ are the ground and the excited states of the ancilla qubit, respectively. The dispersive coupling between the ancilla qubit and the storage and the readout cavities are $\chi_{\mathrm{qs}}/2\pi=1.90$~MHz and $\chi_{\mathrm{qr}}/2\pi=3.65$~MHz. A summary of the device parameters are listed in Table~\ref{Table:DeviceParameters}. The readout fidelities of the ancilla are shown in Table~\ref{Table:readout_fidelity}.

Our experiment is realized in the superconducting quantum system controlled by field programmable gate arrays (FPGA), as shown in Fig.~\ref{fig:exp_diagram}. Different FPGA boards are utilized to control the ancilla qubit (FPGA0), the readout cavity (FPGA1) and the data quantum system (FPGA2), respectively. FPGAs are necessary to realize the quantum channels in the main text based on the adaptive control. The detailed geometry of the device and details of the measurement setup can be found in Ref.~\cite{Hu2019}.

\begin{table}
\centering %
\begin{tabular}{cccccc}
\hline
Description & Term  & Measured value &  \tabularnewline
\hline
ancilla frequency & $\omega_{\mathrm{q}}/2\pi$ & 5.692 GHz &  \tabularnewline
storage cavity frequency & $\omega_{\mathrm{s}}/2\pi$ & 7.634 GHz &  \tabularnewline
readout cavity frequency & $\omega_{\mathrm{r}}/2\pi$ & 8.610 GHz &  \tabularnewline
\hline
ancilla anharmonicity & $K_{\mathrm{q}}/2\pi$ & 232 MHz &  \tabularnewline
self-Kerr of storage cavity & $K_{\mathrm{s}}/2\pi$ & 4.23 kHz &  \tabularnewline
\hline
ancilla-storage coupling & $\chi_{\mathrm{qs}}/2\pi$ & 1.90 MHz &  \tabularnewline
ancilla-readout coupling & $\chi_{\mathrm{qr}}/2\pi$ & 3.65 MHz &  \tabularnewline
\hline
ancilla lifetime & $T_{1\mathrm{q}}$ & 30~$\mu$s &  \tabularnewline
ancilla Ramsey coherence time & $T_{2\mathrm{q}}$ & 40~$\mu$s &  \tabularnewline
storage cavity lifetime (decay rate) & $T_{1\mathrm{s}}$ ($\kappa_\mathrm{s}/2\pi$) & 143~$\mu$s (1.1~kHz) &  \tabularnewline
storage cavity coherence time & $T_{2\mathrm{s}}$ & 252~$\mu$s &  \tabularnewline
readout cavity lifetime (decay rate) & $T_{1\mathrm{r}}$ ($\kappa_\mathrm{r}/2\pi$) & 44~ns (3.62~MHz) &  \tabularnewline
\hline
\end{tabular}\caption{Detailed device parameters.}
\label{Table:DeviceParameters}
\end{table}

\begin{table}
\centering %
\begin{tabular}{cccccc}
\hline
Prepared state~~~~  & Readout fidelity & \tabularnewline
\hline
$\ket{g}$~~~~ & $>0.999$ &  \tabularnewline
$\ket{e}$~~~~ & 0.989 &  \tabularnewline
\hline
\end{tabular}\caption{Readout fidelities of the ancilla qubit. The readout fidelity loss for $\ket{e}$ dominantly comes from the ancilla qubit decay during the measurement time (320~ns).}
\label{Table:readout_fidelity}
\end{table}

\subsection{Experimental procedure}
The open system dynamics could be simulated by Trotterized Liouvillians,
with each Trotter step corresponding to a quantum channel
\begin{equation}
\mathcal{E}_{j}\left(\rho\right)=e^{\mathbb{L}_{j}\Delta t}\rho.
\end{equation}
Then, simulations of the propagators of Liouvillians for the data quantum
system can be realized by constructing a series of gates or quantum channels
acting on the state of the system. The effect is in principle
exactly the same as the target Liouvillians acting on the initial
state.

Therefore, the universal simulator~\cite{Buluta108} of an open system
can be realized in three steps:
\begin{enumerate}
\item Initial state $\ket{\psi(0)}$ preparation. Prepare the initial state
of our data quantum system to study the desired quantum evolution. The preparation is realized with the GRAPE method.
\item Implementation of elementary channels $\left\{ \mathcal{E}_{j}\right\} $
to the system. For our digital quantum simulation of the open system,
each channel $\mathcal{E}_{j}$ is realized digitally via an ancillary
qubit coupling to the data quantum system.
\item Measurement of the controllable quantum system. In current experiments,
the quantum system after each Trotter step is characterized.
\end{enumerate}

\subsection{Experimental sequences}

\begin{figure}
\centering \includegraphics{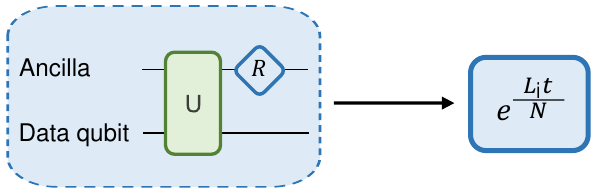}
\caption{Schematic of realizing a quantum channel on a data qubit with the assistance of an ancilla qubit. The quantum channel can be realized by a proper unitary operation on the whole system followed by a reset operation on the ancillary qubit.}
\label{fig:unirtary_reset}
\end{figure}

The data quantum system under study is a bosonic mode in the
storage cavity (the data qubit whose basis states are Fock states \{\ket{0},\ket{1}\}) with an ancillary qubit to mimic the environment. The quantum channel simulation on the data qubit is realized by a unitary gate on the composite data qubit-ancilla system followed by a partial trace of the ancilla, which was
first experimentally realized in Ref.~\cite{Hu2018}. The scheme is schematically shown in Fig.~\ref{fig:unirtary_reset}, where the ancillary qubit is reset after each use of the ancilla. In this work, we implement two different kinds of channels: dephasing channel and amplitude damping channel.

\begin{figure}
\centering \includegraphics[width=8.5cm]{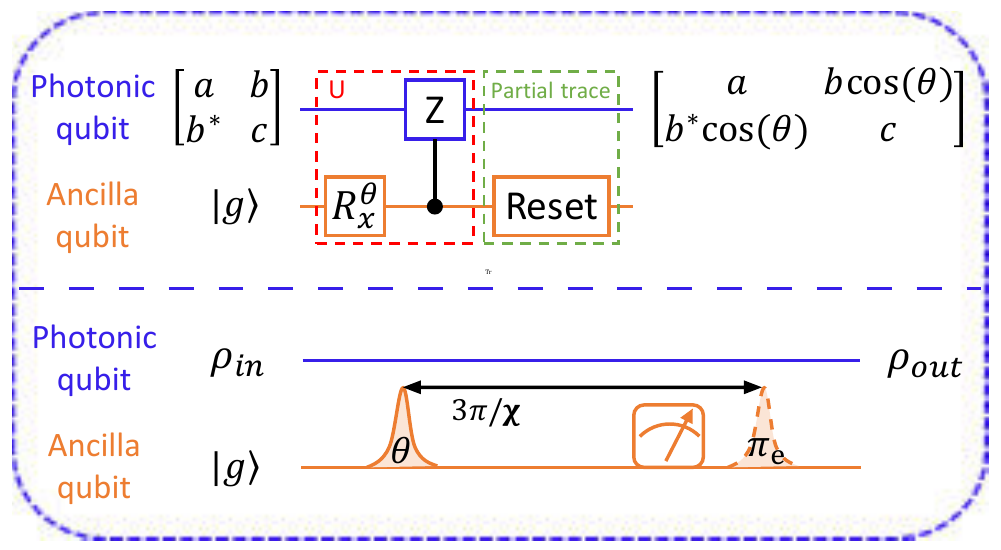} \caption{The circuit and the experimental pulse sequence for realizing the dephasing channel.}
\label{fig:dephasing_channel}
\end{figure}

For the dephasing channel, the circuit and the corresponding experimental
pulse sequence are shown in Fig.~\ref{fig:dephasing_channel}. The realization
can be divided into three steps:
\begin{enumerate}
\item Initialize and prepare the quantum sate of the ancilla qubit by a
rotation gate $R_{x}^{\theta}\left|g\right\rangle $.
\item Realize a controlled-phase (CZ) gate between the ancilla and the data qubit, by utilizing the dispersive interaction ($\chi_{s}\left|e\right\rangle \left\langle e\right|a^{\dagger}a$,
with $\chi_{s}$ being the cross-Kerr coefficient) between the ancilla
qubit and the storage cavity over a time interval of $3\pi/\chi_{s}$.
\item Reset the ancilla to the ground state $\left|g\right\rangle $.
\end{enumerate}

The circuit in the red box in Fig.~\ref{fig:dephasing_channel} represents
the sequence of the unitary gate $U$ on the composite system (including both
the data qubit and the ancilla qubit). The reason for choosing this circuit is that we can adjust the dephasing rate of the channel by changing the rotation angle $\theta$ at will. Ideally, the simulated phase coherence time
is \cite{Hu2018}
\begin{equation}
1/{T_{2}}=-{\rm ln}([2{\rm cos}^{2}(\theta/2)-1])/\tau_{0}+1/{{T_{2}}^{0}}.\label{dampingdephasingidealT2-1}
\end{equation}
Here, $\tau_{0}$ is the time interval for the repetition implementation
of the dephasing channel, and $T_{2}^{0}$ is the intrinsic phase coherence
time of the data qubit. It is worth noting that the unitary gate $U$
could also be realized the GRAPE technique. However, we need
to update the optimized GRAPE pulse when there is a change in the parameter $\theta$,
which would be resource and time consuming.

\begin{figure}
\centering \includegraphics[width=8.5cm]{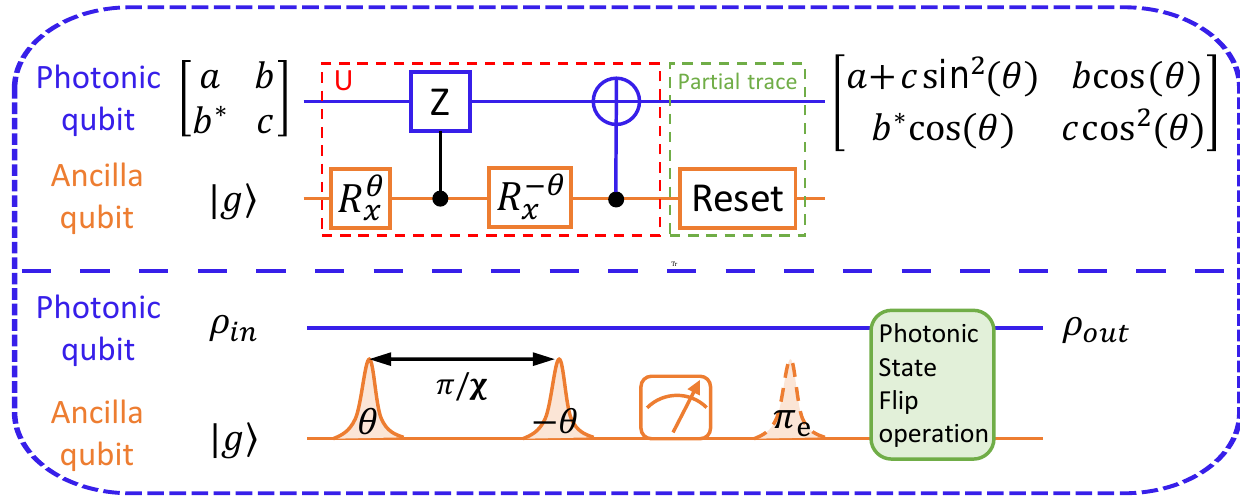} \caption{The circuit and the experimental pulse sequence for realizing the amplitude damping channel.}
\label{fig:damping_channel}
\end{figure}

For the amplitude damping channel, the circuit and the experimental
pulse sequence are shown in Fig.~\ref{fig:damping_channel}. The
unitary gate $U$ is composed by four gates including two ancilla qubit rotation
gates $R_{x}^{\theta}$ and $R_{x}^{-\theta}$, one CZ gate, and one controlled-NOT (CNOT) gate. Similar to the dephasing channel, we can conveniently adjust
the corresponding damping rate of the amplitude damping channel by changing
the angle $\theta$. Here, the CNOT gate is implemented by a measurement-based
adaptive approach~\cite{Hu2018}. Ideally, the circuit gives the simulated coherence
times as
\begin{align}
1/{T_{1}} & =-{\rm ln}({\rm cos}^{2}\theta)/\tau_{0}+1/{{T^{0}_{1}}},\\
1/{T_{2}} & =-{\rm ln}({\rm cos}^{2}\theta)/2\tau_{0}+1/{{T^{0}_{2}}}.
\end{align}
Here $T_1^0$ is the intrinsic energy relaxation time of the data qubit.

The rotation gate on the data qubit is implemented by a GRAPE pulse
acting on the composite system.

\section{Experimental imperfections and calibration}

\subsection{Experimental Imperfections}

In our experimental system, both the ancilla and the data qubit are inevitably coupled to their environment, and consequently the experimental
results would deviate from the above exact formalism of the ideal model.
In our system, we choose the storage cavity with long coherence times
as the data quantum system, while the ancilla is a transmon qubit with
much shorter coherence times. It is cumbersome to do a full analysis of the imperfection completely analytically, so we resort to a numerical model. Based on numerical simulations with the exact experimental pulses and measured device parameters in Table~\ref{Table:DeviceParameters}, our experimental imperfection is mostly attributed to the decoherence of the ancilla. Here, we provide some qualitative analysis of the ancilla error.

For the dephasing channel (Fig.~\ref{fig:dephasing_channel}), there
are two gates: (1) During $R_{x}^{\theta}$, the dephasing and
decay of the ancilla can induce a population uncertainty of the
ancilla, which consequently brings an uncertainty to the phase accumulation of the data qubit. (2) For the CZ gate, the decay of the ancilla can also induce an uncertainty of the accumulated phase of the data qubit, while the dephasing of ancilla has no effect on the data qubit. Therefore, the overall effect of the circuit is to induce a random phase flip on the data qubit. In an ideal circuit, the data qubit accumulates a phase of $\pi$ with a probability of $p=\sin^{2}\frac{\theta}{2}$. 

The situation is more complicated for the amplitude damping channel.
Here, we take the decay of the ancilla as the major error. In Fig.~\ref{fig:damping_channel}, the ancilla decay error mainly appears in the CZ gate and the adaptive CNOT gate. In the CZ gate, the decay could be treated as a reset of the ancilla. When the decay happens, this part is equivalent to
a dephasing channel (Fig.~\ref{fig:dephasing_channel}). The consequent
adaptive CNOT gate is equivalent to a bit-flip channel. So, the decay
error occurs during the CZ gate gives effective both dephasing and bit-flip
channels on the data qubit. Approximately, the decay probability during
the CZ gate is $p\approx0.01\sin^{2}\theta$. In the CNOT gate,
the decay leads to a measurement error of the ancilla and then a
wrong feedforward gate. In the ideal circuit, the data qubit is flipped
from $\left|1\right\rangle $ to $\left|0\right\rangle $ with a probability
of $p=\sin^{2}\theta$, which is determined by the ancilla's population
at $\left|e\right\rangle $ after the $R_{x}^{-\theta}$ gate. However,
the decay of the ancilla would modify this population, and thus reduces the
effective amplitude damping rate. To summarize these two effects, the ancilla
decay error mainly modifies the effective amplitude damping rate of
the target amplitude damping channel and also introduces extra dephasing
and bit-flip channels with a rate $\sim0.01\sin^{2}\theta/\tau_{0}$.
For $\theta\ll1$, we could estimate the extra channel rate would
be about two orders smaller than the target dephasing or damping rate
as $0.01\sin^{2}\theta/\tau_{0}\ll-{\rm ln}({\rm cos}^{2}\theta)/\tau_{0}+1/T_{1}^{0},-{\rm ln}([2{\rm cos}^{2}(\theta/2)-1])/\tau_{0}$.

Other than the above gates, there are also unitary gates on the composite
system by the optimized GRAPE pulses: one is the adaptive gate
on the data qubit and the other one is the coherent Rabi rotation
gate. According to previous demonstrations of the GRAPE gates in our system,
the imperfections of the ancilla and cavity could give rise to a depolarization-like channel to the data qubit with a probability $p_{\mathrm{GRAPE}}=1\%\sim2\%$ for each gate, i.e. the density matrix decays to the identity matrix with a rate of $-\frac{1}{\tau_{0}}\ln\left(1-p_{\mathrm{GRAPE}}\right)$.
Therefore, the error due to the GRAPE unitary gate leads to considerable
errors to the simulation of open quantum system dynamics. Different
from the imperfections in other gates, which contribute small modifications
of the target decoherence rates, the error due to the GRAPE pulse increases with
$1/\tau_{0}$. In order to suppress the influence of this type of
error, we would like to have $-\frac{1}{\tau_{0}}\ln(1-p_{\mathrm{GRAPE}})<\frac{1}{T_{1}^{0}},\frac{1}{T_{2}^{0}}$, and thus we choose $\tau_{0}=3.56~\mathrm{\mu s}$ in our experiments ($\frac{1}{-\frac{1}{\tau_{0}}\ln(1-p_{\mathrm{GRAPE}})}=170\sim350~\mathrm{\mu s}$).

\subsection{Experimental results for calibration}

\begin{figure}[b]
\includegraphics{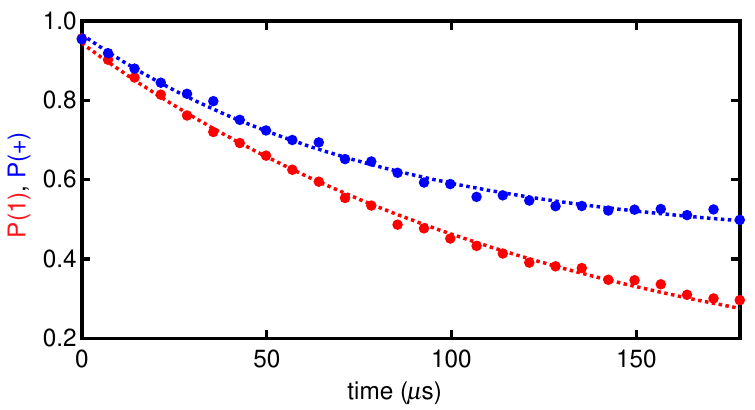} \caption{Experimental results of open quantum system dynamics for calibration, with the parameters $\theta_{1}=\theta_{2}=0$ (parameters in Fig.~2a in the main text). Dots are the experimental data and dashed lines are the fitting curves. Red: Initialize the data qubit to $\ket{1}$, then implement the Trotter process, and finally measure the data qubit on the basis of $\ket{1}$. Blue: Initialize the data qubit to $\ket{+}=(\ket{0}+\ket{1})/\sqrt{2}$, then implement the Trotter process, and finally measure the data qubit on the basis of $\ket{+}$. The effective intrinsic coherence time is obtained by a global fitting of all twelve curves. Here we just present two of the curves.}
\label{fig:op_exp1} \vspace{-6pt}
\end{figure}

As mentioned in the main text, to evaluate the simulation on the open quantum system dynamics we initialize the data qubit $\rho_{in}$ into $\ket{0}$, $\left(\ket{0}+\ket{1}\right)/\sqrt{2}$,
$\left(\ket{0}+i\ket{1}\right)/\sqrt{2}$, and $\ket{1}$ separately,
and measure the corresponding evolution curves of these states on
the bases of the Pauli operators $\sigma_{x}$, $\sigma_{y}$, and
$\sigma_{z}$. Then, by a global fitting of the twelve evolution curves,
we obtain all experimental $T_{1}$, $T_{2}$, and $\Omega$.

To calibrate the influence of the experimental imperfections on the Trotter
simulation of the open quantum system dynamics, we first numerically
implement the Trotter process with $0^{\circ}$ for both dephasing angle
$\theta_{1}$ and damping angle $\theta_{2}$ (Fig.~2a in the main text), which theoretically induces no effect on the coherence time of the data qubit. The representative experimental results are shown in Fig.~\ref{fig:op_exp1}. By a global fitting of the twelve evolution curves, we can get the effective intrinsic coherence times of the data qubit as $T_{1,\mathrm{eff}}^{0}=114~\mathrm{\mu s}$,
$T_{2,\mathrm{eff}}^{0}=80~\mathrm{\mu s}$. These values are much lower than the intrinsic ones, mainly due to the decoherence effects of the ancilla in the numerical model. Therefore, the coherence times in practical experimental system should be modified as
\begin{align}
1/{T_{1}}= & -{\rm ln}({\rm cos}^{2}\theta_{2})/\tau_{0}+1/{{T^{0}_{1,\mathrm{eff}}}},\\
1/{T_{2}}= & -{\rm ln}[2{\rm cos}^{2}(\theta_{1}/2)-1]/\tau_{0}\nonumber \\
 & -{\rm ln}({\rm cos}^{2}\theta_{2})/2\tau_{0}+1/{{T^{0}_{2,\mathrm{eff}}}}.
\end{align}

\begin{figure}
\includegraphics{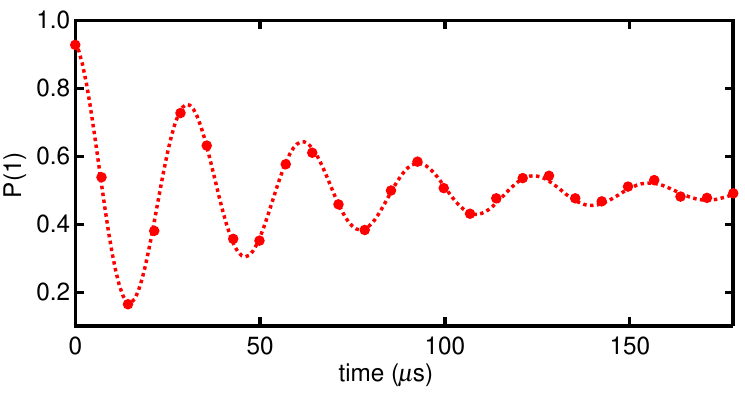} \caption{Experimental results of the open quantum system dynamics with $\theta_{1}=\theta_{2}=0$ and $\theta_{3}=38.6^{\circ}$ (parameters in Fig.~2a in the main text). Dots are the experimental data and dashed line represents the fitting curve. we initialize the data qubit to $\ket{1}$, then implement the Trotter process, and finally measure the data qubit on the basis of $\ket{1}$. The effective intrinsic coherence time is obtained by a global fitting of all twelve curves. Here we just present one
of the curves.}
\label{fig:op_exp2} \vspace{-6pt}
\end{figure}

The effect of the rotation gate is not included in the above numerical simulations. So, we further execute the permutation Dephasing-Damping-Rotating ($\mathcal{L}_{\mathrm{dph}}$-$\mathcal{L}_{\mathrm{damp}}$-$\mathfrak{L}_{\mathrm{R}}$)
channels as an example. We initialize the data qubit to $\ket{1}$, and then
implement the Trotter process with $\theta_{1}=\theta_{2}=0$ and
$\theta_{3}=38.6^{\circ}$. Fitting the curves with the master equation, we get the effective $T^0_{1,\mathrm{eff}}=96~\mathrm{\mu s}$, $T^0_{2,\mathrm{eff}}=82~\mathrm{\mu s}$, and $\Omega=30.2~\mathrm{kHz}$. By comparing $T^0_{1,\mathrm{eff}}$ with and without
the rotation gate, we could evaluate the error of the rotation GRAPE gate
is $\sim1.6\%$, consistent with our estimation. We also find that
the effective $\Omega$ is very close to the target $\Omega$, so we just consider the modifications of $T_{1,\mathrm{eff}}^{0}$ and $T_{2,\mathrm{eff}}^{0}$ in the master equation for the target evolution in the main text.

\section{Numerical simulation of Trotter with different permutations}
\begin{figure}
\includegraphics{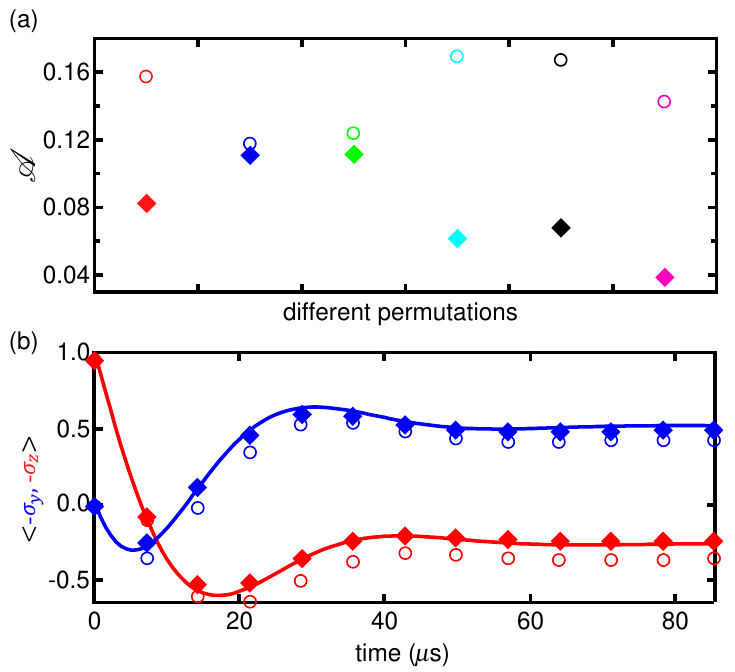} \caption{Numerical QuTip simulation of the 1st-order Trotter and the 2nd-order Trotter. (a) The accuracy of the 1st-order Trotter (hollow dots) and the 2nd-order Trotter (solid diamonds) with dephasing angle $\theta_{1}=20^{\circ}$, damping angle $\theta_{2}=30^{\circ}$, and rotating angle $\theta_{3}=25.7^{\circ}$. Different colors represent different permutations in the form of A-B-C. From left to right, Red: $\mathcal{L}_{\mathrm{damp}}$-$\mathcal{L}_{\mathrm{dph}}$-$\mathfrak{L}_{\mathrm{R}}$; Blue: $\mathcal{L}_{\mathrm{dph}}$-$\mathcal{L}_{\mathrm{damp}}$-$\mathfrak{L}_{\mathrm{R}}$; Green: $\mathcal{L}_{\mathrm{damp}}$-$\mathfrak{L}_{\mathrm{R}}$-$\mathcal{L}_{\mathrm{dph}}$; Cyan: $\mathcal{L}_{\mathrm{dph}}$-$\mathfrak{L}_{\mathrm{R}}$-$\mathcal{L}_{\mathrm{damp}}$; Black: $\mathfrak{L}_{\mathrm{R}}$-$\mathcal{L}_{\mathrm{damp}}$-$\mathcal{L}_{\mathrm{dph}}$; Magenta: $\mathfrak{L}_{\mathrm{R}}$-$\mathcal{L}_{\mathrm{dph}}$-$\mathcal{L}_{\mathrm{damp}}$. (b) Evolution curves of $\sigma_{y}$ (blue) and $\sigma_{z}$ (red) with $\theta_{1}=20^{\circ}$, $\theta_{2}=30^{\circ}$, $\theta_{3}=25.7^{\circ}$. The solid lines are the target evolution curves, solid dots represent the 2nd-order Trotter, and hollow dots represent the 1st-order Trotter. The 2nd-order Trotter results are closer to the target evolution.}
\label{fig:qutip_simulation} \vspace{-6pt}
\end{figure}

To be more realistic, we simulate the experimental sequence numerically in QuTiP~\cite{Johansson2012,Johansson2013} with the experimental pulses and the intrinsic coherence times of the data qubit. Figure~\ref{fig:qutip_simulation}(a) shows the accuracy of the 1st-order Trotter (hollow dots) and the 2nd-order Trotter (solid diamonds), with the dephasing angle $\theta_{1}=20^{\circ}$, damping angle $\theta_{2}=30^{\circ}$, and rotating angle $\theta_{3}=25.7^{\circ}$. Different colors represent different permutations. One can see that every permutation of the 2nd-order Trotter performs better than all the permutations of the 1st-order Trotter. To compare the accuracy of these two kinds of Trotter experimentally, we choose the most accurate permutation of the 1st-order Trotter ($\mathcal{L}_{\mathrm{dph}}$-$\mathcal{L}_{\mathrm{damp}}$-$\mathfrak{L}_{\mathrm{R}}$) and the most accurate permutation
of the 2nd-order Trotter ($\mathfrak{L}_{\mathrm{R}}$-$\mathcal{L}_{\mathrm{dph}}$-$\mathcal{L}_{\mathrm{damp}}$). Figure~\ref{fig:qutip_simulation}(b) shows the evolution curves of the QuTiP simulation on the bases of $\sigma_{y}$ and $\sigma_{z}$, from which we find that the 2nd-order Trotter provides evolution curves nearly the same as the target ones, and is obviously more accurate than the 1st-order Trotter.

\section{Error mitigation}
In order to achieve high-precision quantum computation with a noisy intermediate-scale quantum machine, a new technique called error mitigation~\cite{Temme2016Error,Kandala2019Error}
was proposed for certain tasks, which allows the inference of ideal results, without noise, by extrapolating data from several experimental noise intensities. The value of a physical quantity $E(\lambda)$ in the experiment with a noise intensity $\lambda$ can be represented by the Taylor's expansion around the zero-noise
value $E^{\star}$:
\begin{equation}
E(\lambda)=E^{\star}+\sum_{k=1}^{n}a_{k}\lambda^{k}+\mathcal{O}(\lambda^{n+1}).
\end{equation}
Suppose we can obtain the experimental results $E(\lambda_i)$ with a series of noise intensities
\begin{equation}
\lambda_{i}=c_{i}\lambda,
\end{equation}
with $c_{i}$ ($c_{0}=1$) being a dimensionless scaling factor. Then, we can estimate $E^{\star}$ as a linear combination of the measured $E(\lambda_i)$ with a precision to the $n$-th order by Richardson's deferred
approach to the limit~\cite{Richardson1927The,Sidi2003Practical}:
\begin{align}
E^{\star} &=E_n^{\star}+\mathcal{O}(\lambda^{n+1})\\ &=\sum_{i=0}^{n}\gamma_{i}E(\lambda_i)+\mathcal{O}(\lambda^{n+1}),
\label{eq: extrapolation}
\end{align}
where the coefficients $\gamma_i$ are functions of $c_i$'s. This is achieved by solving the following $n+1$ equations:
\begin{equation}
E(\lambda_i)=E_n^{\star}+\sum_{k=1}^{n}a_{k}(\lambda_i)^{k},
\end{equation}
where $i=0,1,\cdots,n$ and $E_n^{\star}$ is the $n$-th order estimation of the expectation value.

We do the first three orders of Richardson extrapolations with the experimental data sets in Table~\ref{Table: extrapolation}. The error bars in the inset of Fig.~3(d) in the main text are the statistical standard deviations calculated through propagation of the errors of $E(\lambda_i)$ in the estimation relation $E_n^{\star}=\sum_{i=0}^{n}\gamma_{i}E(\lambda_i)$.

\begin{table}[h]
	\caption{The data sets used to do Richardson extrapolation.}\label{Table: extrapolation}\centering
	\setlength{\tabcolsep}{5mm}
	\begin{tabular}{cccc}
		\hline
		\hline
		$i$ & $1/T_1 ~ (\mu s^{-1})$ & $c_i$	& $T_2^\star$~($\mu$s) \\
		\hline
		0 & 0.0090 &1	    & 35.56 \\
		1 & 0.0190 &2.13	& 29.63 \\
		2 & 0.0442 &4.93	& 22.00 \\
		3 & 0.0892 &9.96	& 14.15 \\
		\hline
		\hline
	\end{tabular}
\end{table}


%